# PHYSICAL MODELING AND SIMULATION OF THERMAL HEATING IN VERTICAL INTEGRATED CIRCUITS

Abderrazzak El Boukili

School of Science and Engineering, Al Akhawayn University, Ifrane, Morocco

a.elboukili@aui.ma

## ABSTRACT

Interconnect is one of the main performance determinant of modern integrated circuits (ICs). The new technology of vertical ICs places circuit blocks in the vertical dimension in addition to the conventional horizontal plane. Compared to the planar ICs, vertical ICs have shorter latencies as well as lower power consumption due to shorter wires. This also increases speed, improves performances and adds to ICs density. The benefits of vertical ICs increase as we stack more dies, due to successive reductions in wire lengths. However, as we stack more dies, the lattice self-heating becomes a challenging and critical issue due to the difficulty in cooling down the layers away from the heat sink. In this paper, we provide a quantitative electro-thermal analysis of the temperature rise due to stacking. Mathematical models based on steady state non-isothermal drift-diffusion transport equations coupled to heat flow equation are used. These physically based models and the different heat sources in semiconductor devices will be presented and discussed. Three dimensional numerical results did show that, compared to the planar ICs, the vertical ICs with 2-die technology increase the maximum temperature by 17 Kelvin in the die away from the heat sink. These numerical results will also be presented and analyzed for a typical 2-die structure of complementary metal oxide semiconductor (CMOS) transistors.

## 1. INTRODUCTION

Modern technology of vertical ICs stacks active layers of transistors one above the other separated by insulating oxide, and connected to each other by metal wires. Vertical ICs are also called three dimensional integrated circuits (3D ICs). This technology has the advantage of reducing significantly wire lengths, increasing speed, and providing lower power consumption. However, as we stack more transistors, the power density increases causing the temperatures to increase mainly in the transistors away from the heat sink [1]. And it is well known that self-heating limits the performance of semiconductor electronic and optoelectronic devices as high power laser diodes, high power transistors, high electron mobility transistors (HEMTs), or CMOS transistors [2]. Consequently, the self-heating will also limit the performances of the 3D ICs technology. Heat is generated in semiconductor devices when carriers (electrons and holes) transfer part of their energy to the crystal lattice. Then, the thermal (vibrational) energy of the lattice rises, which is measured as an increase in its temperature, $T_L$. Within the semiconductor lattice, the energy is dissipated by traveling lattice vibrations. The smallest energy portions of lattice waves are called phonons, which can be treated as particles. Microscopic theories of lattice heat generation and dissipation are based on phonons.

The energy transfer from carriers to lattice can occur by diffusion, convection, or radiation. This will depend on the semiconductor device under hand. For physical and mathematical modeling issues, we assume that the heat transfer from carriers to lattice, in CMOS devices, is due to diffusion only. For these CMOS devices, we also assume a local thermal equilibrium between lattice and carriers. Then, the lattice temperature $T_L$ is considered to be the same as the electrons and holes temperatures $T_n$, and $T_p$, respectively. For these reasons, the steady state non-isothermal drift-diffusion model which involves only diffusion terms is enough for our simulations. For other devices, as HEMTs or laser diodes, the energy transfer from carriers to lattice may occur by diffusion, convection or even radiation. Then, a transient or steady state non-isothermal Energy Balance or Hydrodynamic models should be used.

The main focus of this paper is the 3D modeling and simulation of electro-thermal self-heating of 3D ICs with two active CMOS layers as shown in Figure 1. The materials used in this 3D ICs are: Aluminum (Al), Polysilicon (Poly), Silicon Dioxide (SiO2), and Silicon (Si). The mathematical model used is based on steady state non-isothermal drift-diffusion model which involves Poisson's equation and electrons and holes transport equations coupled to heat flow equation for lattice temperature. These models implement the Wachutka's thermodynamically rigorous model of lattice heating [3].

Almost all of the mathematical models used for thermal analysis, by many researchers and found in the literature, are only solving the heat equation as in [4]-[17]. And the heat source in this heat equation is assumed to be given. I would say that this is a too simplified model. In our case, we are using an accurate and comprehensive mathematical model that couples the heat equation to the electrical non-isothermal drift-diffusion equations. In our model, the heat sources are modeled accurately and properly and are depending strong on electrical currents and lattice temperature.

Heat is generated in semiconductor lattice whenever physical processes transfer energy to the crystal lattice. Depending on different energy transfer mechanisms, heat sources can be separated into: Joule heat, electron-hole (radiative and nonradiative) recombination heat, electron-hole generation cooling, Thomson heat, Peltier heat, and optical absorption heat. The mathematical models of these different heat sources will be reviewed and discussed. The models of lattice heat sources that we have used will be presented.

This paper is organized as follows. Section 2 outlines the physically based steady state non-isothermal drift-diffusion and lattice heat flow equations. This section presents and discusses different models for lattice heat and cooling sources in semiconductor devices. It does also outline the numerical methods used to find the lattice temperature distribution in a typical 3D ICs structure. Section 3 presents the computational methods and algorithms used to solve and decouple the equations. Section 4 presents the 3D numerical results and analysis for the 3D ICS structure given in Figure 1. Section 5 discusses the qualitative and quantitative validation of the simulation results. This section also gives a comparison between our results and other results found in the literature [5],[6],[7],[8],[9]. Section 6 holds the concluding thoughts.

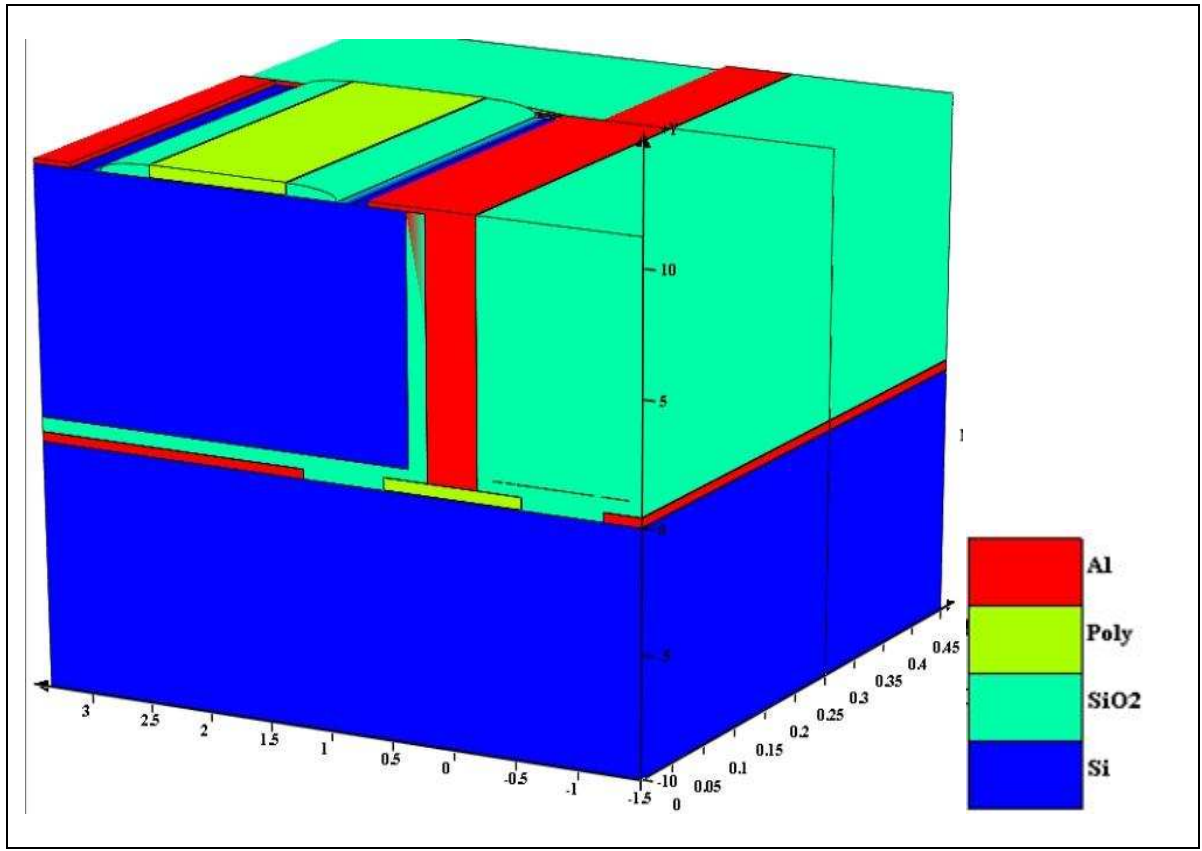

Figure 1. 3D ICs with 2 stacked active CMOS layers.

## 2. PHYSICALLY BASED MATHEMATICAL MODELS

Mathematical models of the operation and fabrication of any semiconductor device result from many years of academic and industrial research into process and device physics. The accuracy of the numerical simulation results depend strongly on the accuracy of the physically based mathematical models. In this paper, we are doing our best to use accurate physically based models.

2.1 Non-Isothermal Drift-Diffusion Model
The non-isothermal drift-diffusion model which takes into account the lattice self-heating effects consists of a set of fundamental equations which link together the lattice local temperature $T_L$, electrostatic potential $\phi$, and the quasi-Fermi levels $\phi_n, \phi_p$ for electrons and holes, respectively. These equations, which are solved inside any general purpose device simulator, have been derived from Maxwell's laws or from semiconductor Boltzmann equations [10]. They consist of Poisson's equation and the transport equations for electrons and holes. Poisson's equation relates variations in electrostatic potential to local charge densities. The transport equations describe the way that the electron and hole densities evolve as a result of transport processes, generation processes, and recombination processes. In the steady state case, these 3 equations are defined as follows [10].

$$div(\varepsilon \nabla \phi) - q(N(\phi,\phi_n,T_L) - P(\phi,\phi_p,T_L) - D) = 0 \quad (1)$$

$$div(J_n) - qGRn(\phi,\phi_n,\phi_p,T_L) = 0 \quad (2)$$

$$div(J_p) + qGRp(\phi,\phi_n,\phi_p,T_L) = 0 \quad (3)$$

The equation (1) represents the Poisson's equation. The equations (2) and (3) represent the transport equations for electrons and holes, respectively. The term $\varepsilon$ represents the local permittivity, $q$ is the magnitude of the charge of an electron. The electron and hole densities $N(\phi,\phi_n,T_L)$ and $P(\phi,\phi_p,T_L)$, respectively, represent the mobile charges. In Boltzmann statistics, they are given by:

$$N(\phi,\phi_n,T_L) = Nc(T_L)\exp(\frac{q(\phi-\phi_n)}{k_B T_L}) \quad (4)$$

$$P(\phi,\phi_p,T_L) = Nv(T_L)\exp(\frac{-q(\phi-\phi_p)}{k_B T_L}) \quad (5)$$

where $Nc(T_L), Nv(T_L), k_B$ represent the effective density of states for electrons and holes and the Boltzmann constant, respectively. In Poisson's equation, D(x) represents the fixed and ionized charges [11]. The non-isothermal carrier current densities $J_n$ and $J_p$ that account for spatially varying lattice temperature are given by:

$$J_n = -q\mu_n N(\phi,\phi_n,T_L)(\nabla\phi_n + P_n \nabla T_L) \quad (6)$$

$$J_p = -q\mu_p N(\phi,\phi_p,T_L)(\nabla\phi_p + P_p \nabla T_L) \quad (7)$$

where $\mu_n$ and $\mu_p$ represent electron and hole mobilities which may depend on lattice temperature $T_L$ and on electric field. These carrier mobilities are the key material parameter in transport simulations. They are limited by collisions of electrons and holes with other carriers, with crystal defects, and with phonons (lattice vibrations). Those scattering events slow down the carriers and constitute the electrical resistance of the material. Various and advanced models for carrier mobilities could be found in [2]. Lattice temperature variations is an additional driving force for thermal current. The generation of current by temperature gradient $\nabla T_L$ is called the Seebeck effect with the thermoelectric powers $P_n$ and $P_p$ $(V/K)$, respectively, as material parameter. The thermoelectric powers account for the extra energy of carriers above the Fermi level. This energy increases with higher temperature due the wider spreading of the Fermi function. When a temperature gradient occurs, carriers move from hot regions to cold regions in order to reduce that extra energy. For the non degenerate semiconductors, $P_n$ and $P_p$ are given by:

$$P_n = \frac{k_B}{q}\left[\ln(\frac{N(\phi,\phi_n,T_L)}{Nc}) - (\frac{5}{2}+ksn)\right] \qquad (8)$$

$$P_p = \frac{k_B}{q}\left[\ln(\frac{P(\phi,\phi_p,T_L)}{Nv}) - (\frac{5}{2}+ksp)\right] \qquad (9)$$

where the values of $ksn = ksp$ depend on the dominant carrier scatter mechanisms [12].

$\frac{5}{2}+ksn = 1, 2, 3, 4, 2.5$ for amorphous semiconductors, for acoustic phonon scattering, for optical phonon scattering, for ionized impurity scattering, and for neutral impurity scattering, respectively. For our simulations, we are taking: $\frac{5}{2}+ksn = 1$. The models for the electron-hole generation and recombination $GRn(\phi,\phi_n,\phi_p,T_L)$ and $GRp(\phi,\phi_n,\phi_p,T_L)$ are detailed in the following section.

## 2.2 Electron-Hole Generation and Recombination Models

The net generation and recombination rates for electron-hole pairs are represented by $GRn(\phi,\phi_n,\phi_p,T_L)$ and $GRp(\phi,\phi_n,\phi_p,T_L)$, respectively. In steady state case, they are given by:

$$GRn(\phi,\phi_n,\phi_p,T_L) = GRp(\phi,\phi_n,\phi_p,T_L) = R(\phi,\phi_n,\phi_p,T_L) - G(\phi,\phi_n,\phi_p,T_L) \qquad (10)$$

In the above equation (10), $R(\phi,\phi_n,\phi_p,T_L)$ represents the total electron-hole recombination rate, and $G(\phi,\phi_n,\phi_p,T_L)$ represents the total electron-hole generation rate. Accurate models for $R(\phi,\phi_n,\phi_p,T_L)$ and $G(\phi,\phi_n,\phi_p,T_L)$ are essential for lattice self-heating simulations as they do represent a source of lattice heating or cooling as we will see later on. The model of $R(\phi,\phi_n,\phi_p,T_L)$ and $G(\phi,\phi_n,\phi_p,T_L)$ depend on the device under hand. For a laser device simulations, all possible electron-hole recombination or generation mechanisms should be included. And, in this case, $R(\phi,\phi_n,\phi_p,T_L)$ is given by:

$$R(\phi,\phi_n,\phi_p,T_L) = R_{SRH} + R_{Auger} + R_{Spont} + R_{Stim} \qquad (11)$$

where $R_{SRH}$ represents the electron-hole recombination due to Shockley-Read-Hall [11]. $R_{SRH}$ involves energy levels deep inside the semiconductor band gap that are generated by crystal defects. Such deep level defects are able to capture electrons from the conduction band as well as holes from the valance band and thereby serve as recombination centers. They are characterized by capture coefficients $c_n$ and $c_p$, trap density $N_t$, and trap energy $E_t$. In the steady state case, $R_{SRH}$ is given by:

$$R_{SRH} = N_t \frac{c_n c_p (NP - N_0 P_0)}{c_n (N + N_1) + c_p (P + P_1)} \tag{12}$$

where $N$ and $P$ represent the electron and hole concentrations defined in the equations (4) and (5). And, $N_1 P_1 = N_0 P_0$ where $N_0$ and $P_0$ represent the electron and hole equilibrium concentrations. $R_{Auger}$ represents the Auger recombination [11]. We should note that $R_{SRH}$ and $R_{Auger}$ represent nonradiative recombination (no emission of photons).

In Auger recombination the excess energy is transferred to another electron within the valence or conduction band. Auger recombination may involve different valence bands and the interaction with phonons. The Auger electron-hole recombination rate is given by:

$$R_{Auger} = (c_n(T_L)N + c_p(T_L)P)(NP - N_0 P_0) \tag{13}$$

where $c_n(T_L)$ and $c_p(T_L)$ are the Auger coefficients and can be found in [11]. $R_{Spont}$ and $R_{Stim}$ represent electron-hole recombination rates due spontaneous and stimulated emissions and their models can be found in [11]. We should also note that $R_{Spont}$ and $R_{Stim}$ represent radiative recombination (emission of photons or light). For the simulation of CMOS transistors as in our case, the radiative recombination rates $R_{Spont}$ and $R_{Stim}$ are included in the total recombination rate $R(\phi, \phi_n, \phi_p, T_L)$.

The generation of electron-hole pairs is looked at as a source of lattice cooling. Since it does absorb some of the lattice energy to generate electron-hole pairs. The generation of electron-hole pairs requires the interaction with other particles. And it is may due to phonons (thermal generation), to photons (optical generation), or to other electrons (generation due to impact ionization) [2].

The net recombination rates given above in equations (12) and (13) already include thermal generation as they vanish under thermal equilibrium, $N_1 P_1 = N_0 P_0$.

In laser diode simulations, the total electron-hole pairs generation $G(\phi, \phi_n, \phi_p, T_L)$ can be defined by:

$$G(\phi, \phi_n, \phi_p, T_L) = G_{Optical} + G_{Impact} + G_{Band-to-Band} \tag{14}$$

where $G_{Optical}$ represents the optical electron-hole pairs generation due to photons absorption. This type of generation is the key physical mechanism in photo detectors and other electro absorption devices. Due to absorption, the light intensity decreases as the light penetrates deeper into the device. If we assume that the optical absorption coefficient $\alpha_0$ is uniform, then the model of the optical absorption is given by:

$$G_{Optical} = \alpha_0 \frac{I_{Opt}(0)}{\hbar \omega} \exp(\alpha_0 z) \tag{15}$$

where $\hbar\omega$ represents photon energy, $\hbar$ represents the reduced Plank constant, $\omega$ represents the angular frequency of the incident radiation, and $I_{Opt}(0)$ represents the optical intensity at the surface and $z$ represents the penetration distance [11]. $G_{Impact}$ represents the electron-hole pairs generation due to impact ionization. Impact ionization is of great importance in devices like avalanche photo detectors. Since these devices use high electric field $F$ and high carrier drift velocities to generate electron-hole pairs. Impact ionization is opposite to the Auger recombination as it absorbs the energy of motion of another electron or hole to generate an electron-hole pair [11]. A typical model of impact ionization is given by [11]:

$$G_{Impact}(F) = \frac{\alpha_n(F)J_n + \alpha_p(F)J_P}{q} \qquad (16)$$

where $\alpha_n(F)$ and $\alpha_p(F)$ represent the ionization coefficients for electrons and holes, respectively. The term $F$ represents the electrical field.

$G_{Band-to-Band}$ represents the electron-hole generation pairs due to band-to-band tunneling. In fact, carriers can be generated without additional energy by band-to-band tunneling with strong electric fields $F > 10^6 V/cm$. The model used is given by [11]:

$$G_{Band-to-Band}(F) = A_{bbt} F^{\gamma bbt} \exp(-\frac{B_{bbt}}{q}) \qquad (17)$$

where the values of $A_{bbt}, \gamma bbt, B_{bbt}$ depend on the material and can be found in [11]. We should note that the electron-hole generation due to band-to-band tunneling is not considered a source of lattice cooling as it does not need additional energy.

## 2.3 Lattice Heat Flow Equation

The physical and mathematical modeling of heat generation and dissipation in semiconductor devices or 3D ICs is extremely challenging. All the material parameters such as carrier mobilities, band gaps, conductivities depend on lattice temperature, $T_L$.

Lattice heat is generated or absorbed whenever physical processes transfer energy to the crystal lattice or absorb energy from the crystal lattice. To account for lattice self-heating effects the non-isothermal drift-diffusion equations (1), (2), and (3) should be solved self-consistently with the lattice heat equation defined as follows:

$$div(k(T_L)\nabla T_L) + H(\phi,\phi_n,\phi_p,T_L) = 0 \tag{18}$$

where $k(T_L)$ represents the thermal conductivity. For steady state simulations, the thermal conductivity $k(T_L)$ is the only parameter of equation (18) that must be specified for each material region in the structure. Thermal conductivity varies as function of lattice temperature. Its model is given by:

$$k(T_L) = \frac{1}{a + bT_L + cT_L^2} \tag{19}$$

where $a, b, c$ are constants for each material [e.g., Sze 1981]. $H(\phi,\phi_n,\phi_p,T_L)$ represents the lattice heat generation or absorption. Its model should take into account all possible sources of lattice heating or cooling. For accurate modeling and simulation of lattice heating or cooling, the model of $H(\phi,\phi_n,\phi_p,T_L)$ should be developed properly and accurately. This will be done in the following section.

2.4 Lattice Heat Generation and Absorption Modeling

According to differences in energy transfer mechanisms, heat generation sources can be separated into: Joule heat, electron-hole recombination heat, Thomson and Peltier heat, and optical absorption heat. And in the same way, the sources of heat absorption which may help in lattice cooling are may be due to electron-hole generation mechanisms.

*Joule heat.* The flow of carriers through a semiconductor is accompanied by frequent carrier scattering by phonons. This leads to a continuing energy loss to the lattice. Carriers move from a higher electrostatic potential to a lower potential, and the corresponding energy difference is typically absorbed by lattice as Joule heat, $H_J$ given by:

$$H_J = \left[\frac{J_n^2}{q\mu_n N(\phi,\phi_n,T_L)} + \frac{J_p^2}{q\mu_p P(\phi,\phi_p,T_L)}\right] \tag{20}$$

$H_J$ is proportional to the electric resistance of the material.

*Recombination heat.* When electron-hole pair recombines, the energy lost is either transferred to a photon (light) and this is known as radiative recombination, or to a phonon (heat) and this is known as nonradiative recombination. The average heat released by electron-hole recombination (or absorbed by electron-hole generation) is proportional to the difference between the quasi-Fermi levels. The amount of heat released and absorbed $H_{RG}$ which models lattice heating (due to recombination) and lattice cooling (due to generation) is given by:

$$H_{RG} = q(R(\phi,\phi_n,\phi_p,T_L) - G(\phi,\phi_n,\phi_p,T_L))\left[\phi_p - \phi_n\right] \tag{21}$$

where $R(\phi,\phi_n,\phi_p,T_L)$ and $G(\phi,\phi_n,\phi_p,T_L)$ are given above by the equations (11) and (14), respectively. For CMOS transistor simulations, as in our case, the recombination model of $R(\phi,\phi_n,\phi_p,T_L)$ given by the equation (11) is reduced to:

$$R(\phi,\phi_n,\phi_p,T_L) = R_{SRH} + R_{Auger} \tag{22}$$

Besides trap recombination $R_{SRH}$, this model also includes the Auger recombination $R_{Auger}$. Since the hot carriers generated during Auger recombination eventually lose their energy to phonons.

For laser diodes, or photo detectors, the spontaneous recombination $R_{Spont}$ and the stimulated recombination $R_{Stim}$ may be included in $R(\phi,\phi_n,\phi_p,T_L)$. On the one hand, most of the photons emitted by spontaneous recombination are absorbed by the semiconductor lattice and eventually converted into heat. Stimulated emission of photons, due to stimulated recombination, also leads to some heat generation as those photons are partially absorbed inside the device.

Electron-hole recombination also causes a cooling of carriers above the Fermi level. This contribution is related to the change in thermoelectric power $P_p$ and $P_n$ of holes and electrons, respectively and it is given by [11]:

$$H_P = q(R - G)T_L(P_p - P_n) \tag{23}$$

*Thomson and Peltier heat.* The thermoelectric power $(V/K)$ is a measure for the increase in average carrier excess energy with increasing temperature. It varies with the density of states, carrier concentration, and temperature. Thomson heat $H_{TP}$ is transferred between carriers and lattice as current flows along a gradient of the thermoelectric power. It is given by:

$$H_{TP} = qT_L(J_n \nabla P_n + J_p \nabla P_p) \tag{24}$$

*Optical absorption heat.* When optical waves penetrate a material, their energy can be partially or fully absorbed. The magnitude and the mechanism of absorption depends on the photon energy $h\gamma$. At low photon energies, the light is directly absorbed by the crystal lattice. At typical energies of photons, absorption by free carriers dominates. This will quickly dissipates the energy to the lattice due to very short intra band scattering times. The optical absorption related heat $H_{Optical}$ could be modeled by:

$$H_{Optical} = \alpha_0 \Phi_{Optical} h\omega \tag{25}$$

Where $h$ represents the Plank constant, $\alpha_0$ is a constant, and $\Phi_{Optical}$ represents the photon flux density and it is given by:

$$\Phi_{Optical} = \frac{I_{Opt}}{\hbar\omega} \tag{26}$$

where $I_{Opt}$ represents the magnitude of the optical current. Then a complete model for heat generation and dissipation $H(\phi, \phi_n, \phi_p, T_L)$ may be given as:

$$H(\phi, \phi_n, \phi_p, T_L) = H_J + H_{RG} + H_P + H_{TP} + H_{Optical} \tag{27}$$

In our case, $H_{Optical}$ is omitted.

## 2. COMPUTATIONAL METHODS AND SOLUTION

We use finite volume method to approximate the strongly coupled and nonlinear equations. We use Newton-Raphson's algorithm to linearize (and decouple) the equations. Different implementations of Newton-Raphson's algorithm have been used. In one implementation, the equations are linearized and kept coupled. In another implementation, the equations are linearized and decoupled. More details about Newton-Raphson's algorithm and its different implementations to solve semiconductor equations can be found in [10] . We use direct methods based on LU factorization [10] or Multi-frontal LU factorization with or without pivoting [13] to solve the arising linear systems.

3D meshing algorithms are based on advanced and robust domain decomposition methods, Delaunay meshing algorithms, and surface re-meshing techniques. Advanced algorithms and techniques have also been developed to merge the mesh of two different dies in 3D ICs to a single mesh. To mesh a chip of 3D ICs, we decompose the whole chip into a certain number of blocks. We mesh separately each block using the appropriate mesh generation tools. Then, we merge the mesh of the different blocks into a single global mesh. The equations are then solved on this global mesh. The mesh is refined locally enough to get accurate solutions. No automatic refinement or mesh adaptation procedure has been used. That is an other complicated issue in 3D ICs. Parallelization of all these techniques and algorithms on multi-core processors could also be used.

The 3D numerical results showing substantial thermal increase in CMOS transistors away from the heat sink will be presented for a typical structure given by Figure 1.

## 4. NUMERICAL RESULTS AND ANALYSIS

The standard process flow of 1 microns technology is used to fabricate each CMOS layer in such a multiple layer structure. In this study, we did consider several kinds of thermal environments: multiple layers (from 1 to 3), layer thicknesses (from 10 to 100 microns per layer) and different kinds of thermal boundary conditions (Dirichlet and Neumann). These include heat sink on top and bottom, heat sink on either top or bottom, and a range of thermal conductance for the contact wires which provide additional cooling for the devices. Drichlet boundary conditions are used on the top and bottom of the 3D ICs presented here. Homogeneous Neumann boundaries conditions are used on the remaining boundaries that are assumed to be adiabatic. For electrical boundaries, we are using Dirichlet boundary conditions on the source, gate, and drain and homogeneous Neumann boundary conditions on the remaining boundaries. Typical results will be presented to demonstrate the 3D current flow and the resulting heating effects. Each active layer is 10 microns thick with heat sink on the top and bottom. So, the bottom device is the one away from the heat sink. And the top device is the closest to the heat sink. We are, then, expecting the temperature to be higher in the bottom device. The applied voltages at the gate and drain of the bottom active transistor are 4 volts, respectively to turn on the CMOS transistor. The thermal conductance of the connecting Aluminum wires have been reduced according to the wire lengths. It is significant that 17 Kelvin increase in the temperature of the bottom active layer have been obtained from the simulation. For investigation and comparison purposes, we have set up a similar structure with increased layer thickness. The wiring thermal conductance's have been increased proportionally. We found an important increase of temperature in a thicker structure.

Some typical results are presented here to demonstrate the 3D current flow and the resulting heating effects.

Figure 1 shows a 2 active CMOS layers configuration in 3D. This is a typical structure in 3D ICs. Each active layer is 10um thick with heat sink on the top and bottom. The applied voltages at the gate and drain of the bottom active transistor are respectively 4 volts to turn on the CMOS device. The thermal conductance of the connecting Aluminum wires have been reduced according to the wire lengths.

Figure 2 shows the electrostatic potential distribution in the cross section of the 3D ICs structure. Figure 3 presents the temperature profile in the cross section of a single active layer corresponding to the bottom layer of Figure 1. Figure 4 shows the temperature distribution in the cross section of the 2-active CMOS layers given in Figure 1.

Figure 4 shows that there is 17 Kelvin increase in the temperature of the bottom active layer compared to the single active layer given in Figure 3. This means that in 3D ICs the temperature will increase significantly in the layers away from the heat sink. Similar results to those given in Figure 4 have been reported in [5],[6], [7],[8],[9].

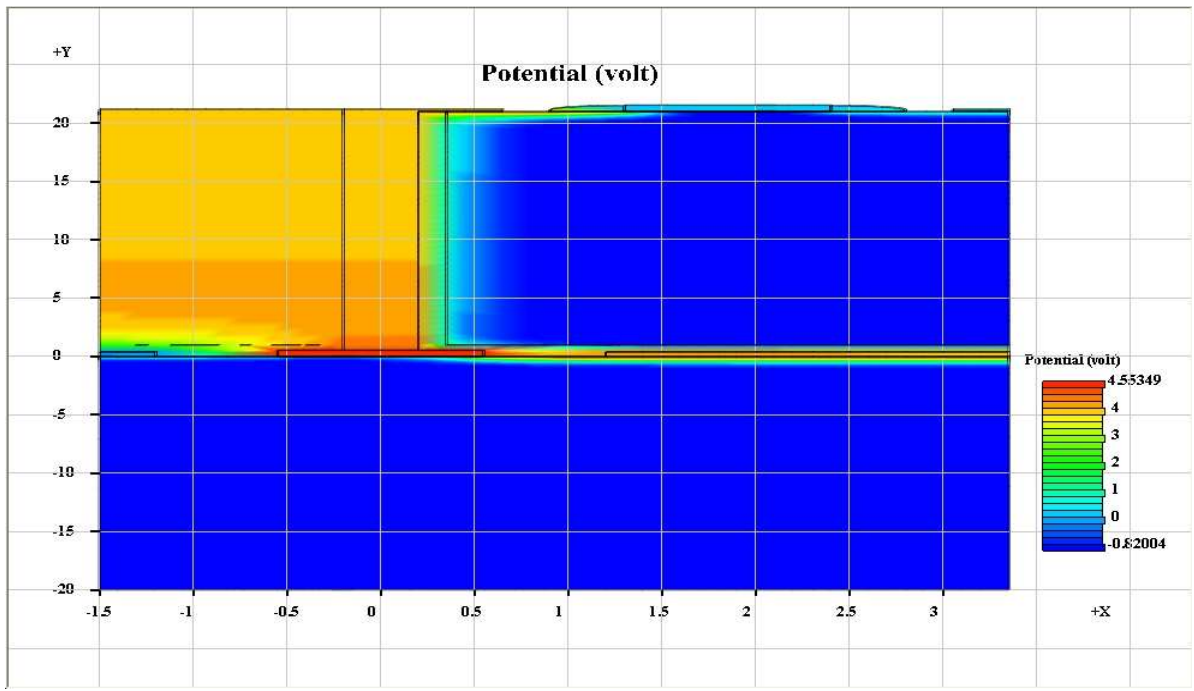

Figure 2.   Potential distribution in the cross section of Figure 1.

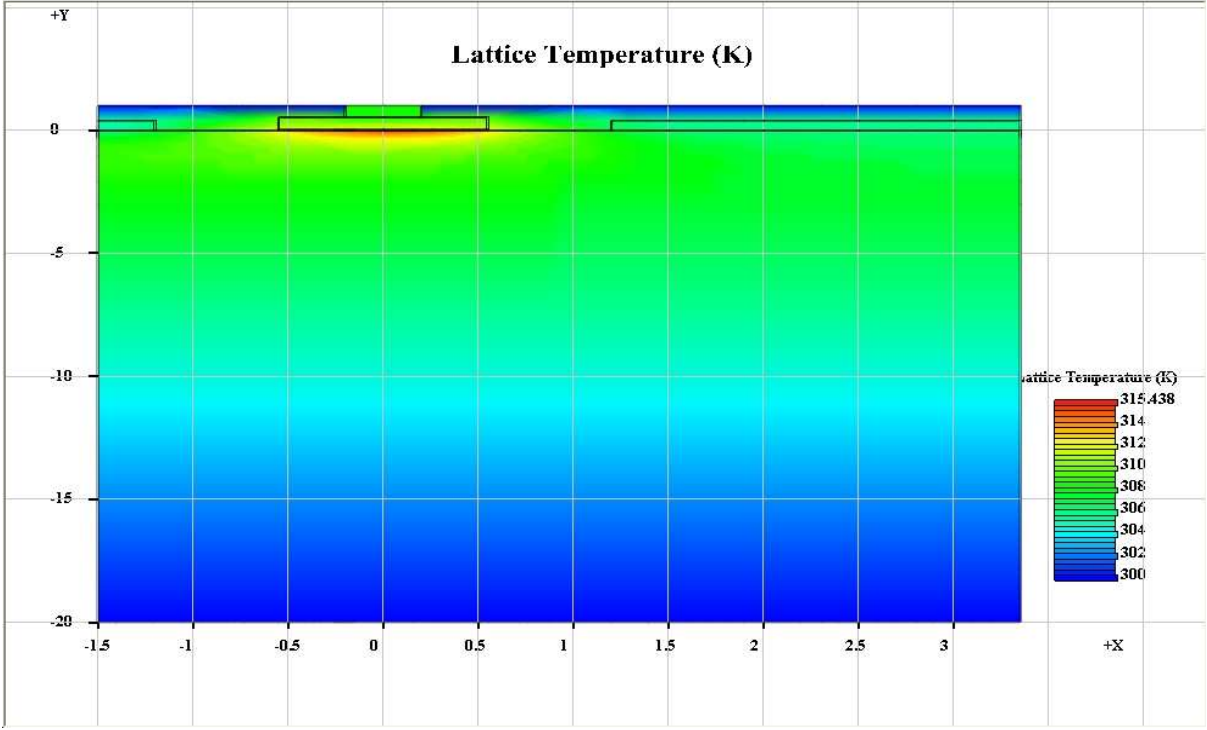

Figure 3.  Temperature distribution in the cross section of a single active CMOS layer.

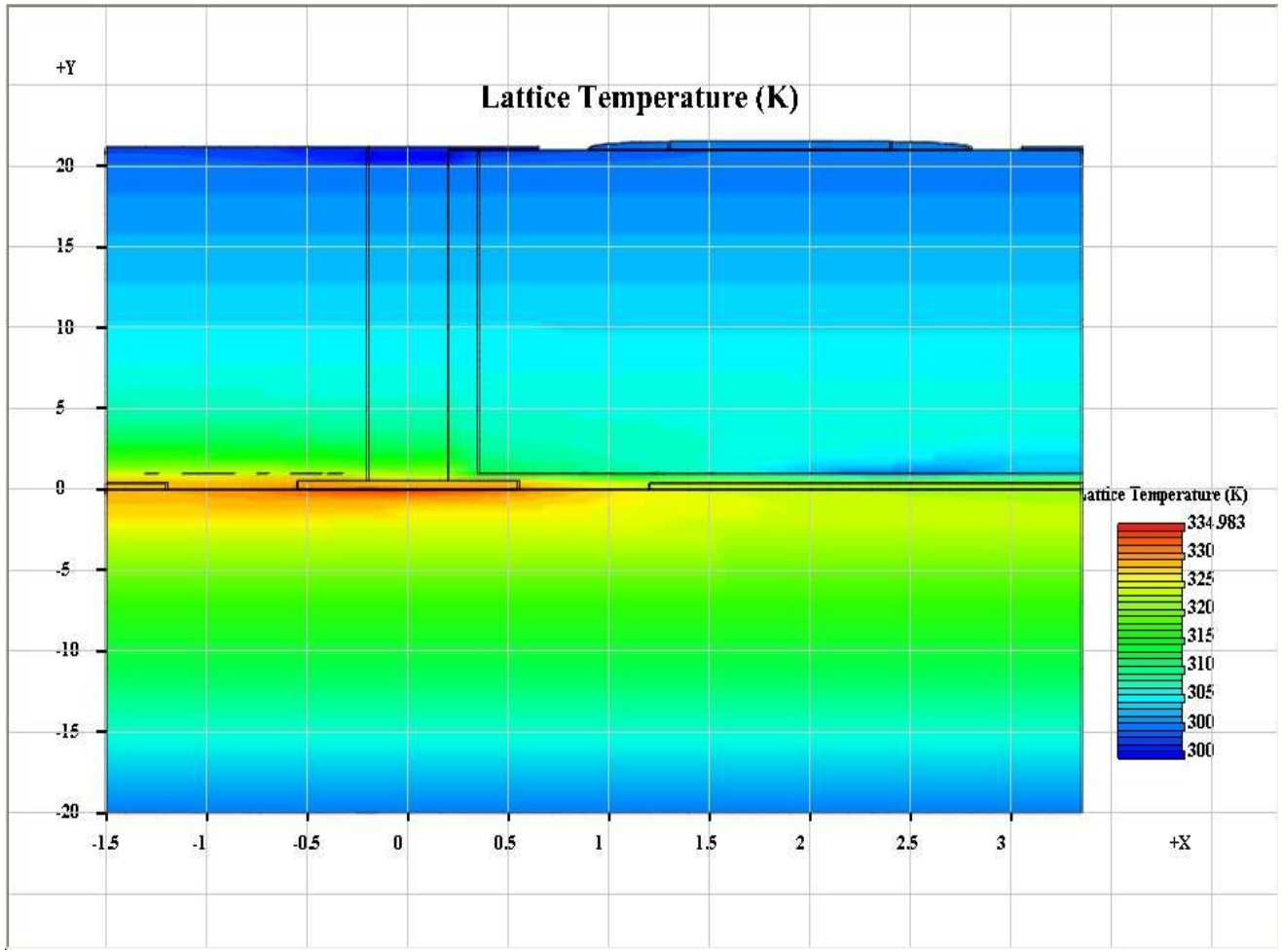

Figure 4. Temperature distribution in the cross section of 2-active CMOS layers of Figure 1.

## 5. Validation and comparison of the simulation results

The result in Figure 4 already proves the validity of our simulation results. Since the result in the Figure 4 shows that there is 17 Kelvin increase of temperature in the bottom device which is the one away from the heat sink. Similar thermal results, for similar 2-die 3D ICs, have been reported in [5],[6], [7],[8],[9]. For example, in [5], the authors analyzed the thermal impact of 3D ICs technology on high-performance microprocessors by computing the temperatures of a planar IC based on the Alpha 21364 processor as well as 2-die and 4-die 3D IC implementations of the same. They have only solved numerically the heat equation where the heat source is given.

The thermal profile of the planar IC in Figure 6 in [5] shows that the maximum temperature is 312 Kelvin. And the thermal profile of the 3D IC with 2-die shown in Figure 7 of [5] shows that the maximum temperature is 328 Kelvin. This means that there is 16 Kelvin increase of temperature in the die away from the heat sink. In our case, we found a 17 Kelvin increase of temperature in the die away from the heat sink as shown in Figure 4. Then our results are quantitatively comparable to those found in [5].

## 6. CONCLUSION

In conclusion, robust meshing algorithms have been used to build successfully a 3D stacked CMOS structure. And the electro-thermal investigation and analysis based on advanced, physically based, mathematical models and numerical simulations did show substantial temperature increase in CMOS devices away from the heat sink. The exact temperature increase due to layer stacking is sensitive to layer thickness and wiring thermal boundary conditions. The new challenges, in 3D ICs, are again making the technology computer aided design simulation tools crucial and mandatory in designing, optimizing and analyzing 3D ICs technology.

## REFERENCES

[1] S. Das, A. Chandrakasan, R. Reif, (2004) "Timing, energy, and thermal performance of three-dimensional integrated circuits". GLSVLSI 04:*14$^{th}$ ACM Great Lakes symposium on VLSI*, pp338-343.

[2] J. Piprek, (2003) *"Semiconductor optoelectronic devices, introduction to physics and simulation"*, Elsevier Science Publ., San Diego, California, USA.

[3] G. K. Wachutka, (1990) "Rigorous thermodynamic treatment of heat generation and conduction in semiconductor transmodeling ", *IEEE Trans*.,CAD-9, pp1141-1149.

[4] W. Huang, K. Sankaranarayana, K. Skadron, R. Ribando, M. Stan,(2007) "Accurate, Pre-RTL temperature –aware design using a parameterized, geometric thermal model ". *Design, Automation, and Test in Europe*.

[5] K. Puttaswamy, G. Loh, (2006) "Thermal analysis of a 3D die-stacked high-performance microprocessor ". *Great Lakes Symposium on VLSI*.

[6] P. Michaud, Y. Sazeides,(2007) "Analytical model of temperature in microprocessors". *Workshop on Modeling, Benchmarking and Simulation*.

[7] Y. Zhan, S. Sapatnekar, (2005) "A high efficiency full-chip thermal simulation algorithm". *International Conference on Computer Aided Design*.

[8] D. Oh, C. Chen, Y. Hu, (2007) "3DFFT: Thermal analysis of non-homogenous IC using 3DFFT Green function method". *International Symposium on Quality Electronic Design*.

[9] H. Clarke, K. Murakami, (2011) "Superposition principle applied to thermal analysis for 3D ICs". *International Conference on Circuits, System and Simulation*.

[10] A. El Boukili, (2005) *"Analyse mathematique et simulation numerique bidimentionnelle des transistors bipolaires a heterojunction par elements finies mixtes"*. P.h.D. Thesis, Paris 6 University, Paris, France.

[11] S. M. Sze, (1981) *"Physics of semiconductor devices"*. John Wiley & Sons Publ., New York, USA.

[12] K. W. Boer, (1990) *"Survey of semiconductor physics"*. Vol. I. Van Nostrand Reinhold Publ., New York, USA.

[13] R. Amestoy, I. S. Duff, (1998) *"Excellent, multifrontal parallel distributed symmetric and unsymmetric solvers, Comput. Methods"*. *Appl. Mech. Eng.*,184, pp501-520.

[14] S. Selberherr, (1991) *"Analysis and simulation of semiconductor devices"*. Springer-Verlag, Wien-New York.

[15] J. Kim, S. Jhang, C. John, (2010) *"Dynamic register-renaming scheme for reducing power-density and temperature"*. *Symposium on Applied Computing*.

[16] S. Melamed, T. Thorolofsson, A. Srinisan, A. Cheng, P. Franzon, R. Davis, (2009) *"Junction-level thermal extraction and simulation of 3D ICs"*. *IEEE International 3D Systems Integration Conference.*

[17] H. Oprins, M. Cupak, G. Van Der Plas, P. Marchal, A. Srinivasan, E. Cheng, (2009) *"Fine-grain thermal modeling of 3D stacked structures"*. *15th International Workshop on Thermal Investigations of ICs and Systems*, October, Leuven, Belgium, pp45-49.

**Authors**

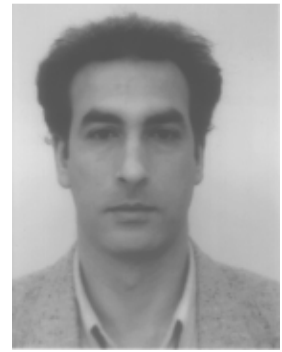

**Abderrazzak El Boukili** received both the PhD degree in Applied Mathematics in 1995, and the MSc degree in Numerical Analysis, Scientific Computing and Nonlinear Analysis in 1991 at Pierre et Marie Curie University in Paris-France. He received the BSc degree in Applied Mathematics and Computer Science at Picardie University in Amiens-France. In 1996 he had an industrial Post-Doctoral position at Thomson-LCR company in Orsay-France where he worked as software engineer on Drift-Diffusion model to simulate heterojunction bipolar transistors for radar applications. In 1997, he had European Post-

Doctoral position at University of Pavia-Italy where he worked as research engineer on software development for simulation and modeling of quantum effects in heterojunction bipolar transistors for mobile phones and high frequency applications. In 2000, he was Assistant Professor and Research Engineer at the University of Ottawa-Canada. Through 2001-2002 he was working at Silvaco Software Inc. in Santa Clara, California-USA as Senior Software Developer on mathematical modeling and simulations of vertical cavity surface emitting lasers. Between 2002-2008, he was working at Crosslight Software Inc. in Vancouver-Canada as Senior Software Developer on 3D Process simulation and Modeling. Since Fall 2008, he is working as Assistant Professor of Applied Mathematics at Al Akhawayn University in Ifrane-Morocco. His main research interests are in industrial TCAD software development for simulations and modeling of opto-electronic devices and processes.

http://www.aui.ma/personal/~A.Elboukili.